\documentclass[%
reprint,
superscriptaddress,
 amsmath,amssymb,
aps,
prb,
]{revtex4-1}

\usepackage[usenames,dvipsnames]{color}

\usepackage{comment}
\usepackage{graphicx}
\usepackage{dcolumn}
\usepackage{bm}

\newcommand{\atan}{\text{atan}}

\begin{document}


\title{Spinon dynamics in quantum integrable antiferromagnets}
\date{\today}
 
\author{R. Vlijm} \email{R.P.Vlijm@uva.nl} \affiliation{Institute for Theoretical Physics, University of Amsterdam, Science Park 904, 1098 XH Amsterdam, The Netherlands}
\author{J.-S. Caux} \affiliation{Institute for Theoretical Physics, University of Amsterdam, Science Park 904, 1098 XH Amsterdam, The Netherlands}

\begin{abstract}
The excitations of the Heisenberg antiferromagnetic spin chain in zero field are known as spinons. As pairwise-created fractionalized excitations, spinons are important in the understanding of inelastic neutron scattering experiments in (quasi) one-dimensional materials. In the present work, we consider the real space-time dynamics of spinons originating from a local spin flip on the antiferromagnetic ground state of the (an)isotropic Heisenberg spin-1/2 model and the Babujan-Takhtajan spin-1 model. By utilizing algebraic Bethe ansatz methods at finite system size to compute the expectation value of the local magnetization and spin-spin correlations, spinons are visualized as propagating domain walls in the antiferromagnetic spin ordering with anisotropy dependent behavior. The spin-spin correlation after the spin flip displays a light cone, satisfying the Lieb-Robinson bound for the propagation of correlations at the spinon velocity. 
\end{abstract}

\maketitle

\section{Introduction} \label{sec:intro}
The existence of fractionalized collective excitations is one of the most interesting features of quantum many-body physics. In antiferromagnetic quantum spin chains, examples of such collective modes are generally known as spinons,\cite{1981_Faddeev_PLA_85} which are fractionalized spin-$1/2$ excitations created pairwise by spin flips on the antiferromagnetic state. A fundamental model containing spinons is the Heisenberg antiferromagnetic spin chain.\cite{1928_Heisenberg_ZP_49} The exact eigenstates are obtained by the Bethe ansatz\cite{1931_Bethe_ZP_71,1958_Orbach_PR_112,KorepinBOOK} as plane waves of magnons with amplitudes derived from their scattering phases. In this context, spinons are the hole-like modes in a sea of interacting magnons at zero field and can be pictured in real space as propagating domain walls in the local antiferromagnetic ordering of the spins. 

Experimentally, spinon physics is demonstrated in inelastic neutron scattering\cite{1991_Nagler_PRB_44,1993_Tennant_PRL_70,1995_Tennant_PRB_52_1,1995_Tennant_PRB_52_2,2004_Zaliznyak_PRL_93,2013_Mourigal_NATPHYS_9,2013_Lake_PRL_111} on quantum spin chains near zero temperature. The incoming neutron interacts with the magnetic moments of the (quasi) one-dimensional spin chain material, without influencing the electronic structure due to its neutral charge. The differential cross section of the scattering neutrons is directly related to the dynamical structure factor,\cite{1954_vanHove_PR_95_1,1954_vanHove_PR_95_2} which is defined as the Fourier transform of the spin-spin correlation
\begin{equation}
S^{a\bar a}(q,\omega)=\frac{1}{N}\sum_{j,j^\prime}^N e^{-i q (j-j^\prime)}\int_{-\infty}^\infty \text{d}t \; e^{i\omega t} \langle S_j^a (t) S_{j^\prime}^{\bar a} (0)\rangle_c . \label{eq:defDSF1}
\end{equation}
The label $a=z,\pm$ distinguishes the longitudinal and transversal structure factors respectively. 
The dynamical structure factor has the shape of the (multi-)spinon continua, while the value of the correlation corresponds to the intensity. The dynamical structure factor therefore serves as an important connection between theory and experiments.

By inserting a resolution of the identity in Eq.~\eqref{eq:defDSF1}, 
the spin-spin correlation becomes a sum over matrix elements of a single spin operator with the ground state and an excited spinon state. This algebraic Bethe ansatz based computation of the dynamical structure factor relies on determinant expressions\cite{1989_Slavnov_TMP_79,1990_Slavnov_TMP_82,1999_Kitanine_NPB_554,2007_Castro_Alvaredo_JPA_40} of matrix elements in terms of the rapidities of the Bethe states. By summing over the matrix elements of spinon states, many properties of the dynamical structure factor have been evaluated for the Heisenberg antiferromagnet at finite system size,\cite{2002_Biegel_EPL_59,2003_Biegel_JPA_36,2004_Sato_JPSJ_73} including the effects of anisotropic interactions\cite{2005_Caux_JSTAT_P09003} and the presence of a magnetic field.\cite{2005_Caux_PRL_95,2009_Kohno_PRL_102}

Analytic results for the dynamical structure factor in the thermodynamic limit exist for the isotropic chain through the vertex operator approach.\cite{JimboBOOK}  Both the two-spinon\cite{1996_Bougourzi_PRB_54,1997_Karbach_PRB_55,1998_Bougourzi_PRB_57,2008_Caux_JSTAT_P08006} and four-spinon\cite{1997_Abada_NPB_497,2006_Caux_JSTAT_P12013} contributions to the dynamical structure factor have been computed by this method, showing that the first moment sum rule is saturated by the two-spinon contributions by $71.3\%$ in the thermodynamic limit, while the four-spinon carries $27(\pm1)\%$. The intensity describing two- and four-spinons in the dynamical structure factor has been observed experimentally \cite{2013_Mourigal_NATPHYS_9} as well.

The aforementioned approaches are mainly focussed on momentum-energy resolved spinon physics, both in experimental and theoretical context. Recent work from Ref.~\onlinecite{2015_Deguchi_ArXiv} explored the real space-time behavior of a linear combination of equally weighted two-spinon states, giving rise to a locally magnetized state diffusing in time. In the present work, we construct the initial state by acting with a local spin flip on the antiferromagnetic ground state, mimicking the spin-flip caused by a neutron scattering off the chain.
This procedure creates precisely the state for which the weights of the spinon states are computed or measured directly from the transverse dynamical structure factor in Eq.~\eqref{eq:defDSF1}.
The time evolved expectation value of the local magnetization $\langle S^z_j (t) \rangle$ can be obtained from algebraic Bethe ansatz techniques at finite size, making it possible to visualize the spinons propagating through the chain.

Additionally, this method allows for the time-evolved evaluation of two-point correlations in real space after the local spin flip. Due to the locality of the interactions, the Lieb-Robinson bound\cite{1972_Lieb_CMP_28} dictates that the buildup of correlations cannot occur faster than the propagation velocity of the spinons. We therefore consider the spin-spin correlation between two sites separated from each other with the spin flip in the center, which become correlated as the spinon passes by on both sites, displaying a light cone around the spin flip. Moreover, we compute the nearest neighbor spin-spin correlation at a fixed site, showing the behavior of the antiferromagnetic correlation during the passing of the spinon. 

This article is structured in the following way. In section~\ref{sec:betheansatz} we elaborate on the construction of spinon states from Bethe ansatz. In section~\ref{sec:trackingspinons}, we show the time evolution magnetization profiles of the spinons, visualizing the propagation of spinons in real space. Section~\ref{sec:BT} focusses on results for the dynamics of two spinons in the Babujan-Takhtajan spin-$1$ chain. Section~\ref{sec:correlations} shows the buildup of correlations while the spinons propagate along the isotropic Heisenberg chain.

\section{Spinons from Bethe ansatz}
\label{sec:betheansatz}
This section reviews the theoretical construction of spinons for the (an)isotropic Heisenberg model from Bethe ansatz. Due to its technical nature it could be skipped on first reading. The spinon states described in this section will be used as a basis for the time evolution after a local spin flip on the ground state, described in section~\ref{sec:trackingspinons}.

The Hamiltonian of the anisotropic Heisenberg model is given as
\cite{1928_Heisenberg_ZP_49,1958_Orbach_PR_112} 
\begin{equation}
  H = J \sum_{j=1}^{N} \left[ S^x_j S^x_{j+1}+S^y_j S^y_{j+1}+\Delta\left(S^z_j
      S^z_{j+1}-\frac{1}{4}\right)\right],
\label{eq:xxzhamiltonian}
\end{equation}
containing an anisotropy in the nearest neighbor interactions given by the parameter $\Delta$. In the present work we restrict to the isotropic XXX model ($\Delta=1$) and the XXZ gapped model ($\Delta>1$). Furthermore, we adapt periodic boundary conditions, identifying $S_{N+1}=S_{1}$.

The fully polarised state $|0\rangle$ acts as the reference state for the Bethe ansatz wave functions. The total spin along the $z$-axis $S^z_{\textrm{tot}}=\sum_{j=1}^N S^z_j$ commutes with Hamiltonian~\eqref{eq:xxzhamiltonian}, splitting the Hilbert space into sectors of fixed magnetization, denoted by the number of down spins $M$ starting from the reference state.

The Bethe ansatz wave functions\cite{1931_Bethe_ZP_71,1958_Orbach_PR_112} for Hamiltonian~\eqref{eq:xxzhamiltonian} are plane waves of magnons
\begin{equation}
	|\{\lambda\}\rangle = \!\!\!\! \sum_{{j_1 < \ldots < j_M}}\!\! \sum_{Q}A_Q(\{\lambda\})\prod_{a=1}^M e^{i j_a p(\lambda_{Q_a})} S_{j_a}^- | 0 \rangle,
	\label{eq:bethestates}
\end{equation}
where the amplitudes $A_Q$ are given by the scattering phases. Most importantly, every wave function in this equation is parameterized in terms of a unique set of non-coinciding rapidities $\{\lambda
\}_M$ obeying Bethe equations, derived from the scattering phases between the magnons and periodic boundary conditions. In logarithmic form the Bethe equations read
\begin{equation}
\theta_1(\lambda_j) - \frac{1}{N} \sum_{k \neq j}^M \theta_2(\lambda_j-\lambda_k) = \frac{2 \pi}{N} J_j,
\label{eq:logbetheeqs}
\end{equation}
where 
\begin{equation}
\theta_n(\lambda)=2\atan(2\lambda/n)
\label{eq:thetadef1}
\end{equation}
for $\Delta=1$ and
\begin{equation}
\theta_n(\lambda)=2\atan\left(\frac{\tan \lambda }{ \tanh(n\zeta /2)}\right)+2\pi \lfloor \frac{\lambda}{\pi} + \frac{1}{2}\rfloor 
\end{equation}
 for $\Delta>1$, with $\zeta=\textrm{acosh} (\Delta)$ and $\lfloor \lambda \rfloor$ being the floor function. The Bethe quantum numbers $J_j$ are (half-odd) integers for $N+M$ (even) odd.

For the isotropic case $\Delta=1$, the model has a global $SU(2)$ symmetry, for which the highest weight states are given by Bethe states with finite rapidities only. Lower weight states are constructed by placing one or more rapidities at infinity, see for example Refs.~\onlinecite{GaudinBOOK} and~\onlinecite{GaudinTRANSLATION}.

The maximum allowed quantum number $J^{\max}$ is determined by sending one of the rapidities to infinity and computing the associated quantum number by means of the Bethe equations. By considering these limiting quantum numbers, the dimensionalities of different classes of excitations can be determined.

The ground state $| \{ \lambda_\text{GS}\}\rangle$ at zero magnetization consists of a set of $M=N/2$ real rapidities, characterized by the set of quantum numbers $\{ J_j \}=\{-\frac{N}{4}+\frac{1}{2}, ..., \frac{N}{4}-\frac{1}{2} \}$. This set of real rapidities is unique, there are no other possibilities to arrange $M=N/2$ real rapidities in this magnetization sector.

The first excitations from the ground state induced by a spin flip are again sets of real rapidities, with $M=N/2-1$. By taking the limit $\lambda \rightarrow \infty$ in the Bethe Eqs.~\eqref{eq:logbetheeqs}, the maximum allowed value of the quantum number $J^{\max}=J^\infty-1=N/4$ is determined. This implies there are $2J^{\max}+1= N/2+1$ possible slots for the quantum numbers, over which $N/2-1$ quantum numbers must be distributed. This set of excitations is therefore characterized by two holes in the sea of quantum numbers, named as two-spinon excitations. The dimension of the set of excitations (for the sector $s=1$, $s^z=0$) created in this way is
\begin{equation}
\dim \mathcal H^{\Delta=1}_{2\textrm{sp}} ={ N/2+1 \choose N/2-1}=\frac{N}{8}(N+2).
\label{eq:dim2sp}
\end{equation}

For the regime with $\Delta>1$, the correct counting of the two-spinon states becomes more complex, as the global symmetry becomes of quantum group nature. Building the two-spinon states from the ground state gives the same set of two-spinon excitations as for the isotropic case. There however exists an additional set of two-spinon excitations,\cite{2008_Caux_JSTAT_P08006} built from the quasi-degenerate ground state with momentum $\pi$, created by an Umklapp from the true ground state, $\{ J_j \}=\{-\frac{N}{4}+\frac{3}{2}, ..., \frac{N}{4}+\frac{1}{2} \}$. Avoiding double counting of the excitations created from the true and quasi-degenerate ground state, the dimension of the two-spinon excitations for $\Delta>1$ is
\begin{equation}
\dim \mathcal{H}^{\Delta>1}_{2\textrm{sp}}=N^2/4.
\label{eq:dim2spxxzgpd}
\end{equation}

At high $\Delta$, the resulting physics after acting with a spin flip operator on the ground state becomes dominated by these two-spinon contributions only. We therefore restrict ourselves to the subspace of two-spinon states in our further analysis for $\Delta>1$. 

Contrary to the limit of high anisotropy, higher spinons contribute for a significant part of the dynamical structure factor at $\Delta=1$. To construct higher spinon states, one has to consider complex solutions of the Bethe equations. The full set of rapidities $\{ \lambda \}$ must remain self-conjugate, and takes the form of the string hypothesis, a string of length $n$ being defined as
\begin{equation}
  \lambda_{j,a}^{(n)}=\lambda_{j}^{(n)}+\frac{i \zeta}{2}(n+1-2a)
  +i \delta^{(n)}_{j,a}\, 
  \label{eq:defstrings}
\end{equation}
with internal label $a = 1, ... , n$. The set of $M$ complex rapidities is now grouped as $M_n$ $n$-strings, where $\sum_n n M_n=M$. At finite system size, the strings are deformed by the deviation $ \delta^{(n)}_{j,a}\in \mathbb{C}$, which in general is exponentially small in system size. There exist however many exceptions with large deviations. Nonetheless, for the classification of the string configurations, we can adopt the limit of vanishing string deviations to derive the Bethe-Gaudin-Takahashi equations\cite{1972_Takahashi_PTP_48} in terms of the $n$-string centers~$\lambda_j^{(n)}$,
\begin{multline}
  \theta_n(\lambda_j^{(n)})-\frac{1}{N} \sum_m \sum_{k=1}^{M_m} \Theta_{nm}(\lambda_j^{(n)}-\lambda_k^{(m)})=\frac{2\pi}{N}I_j^{(n)}\,,\\ 
j=1,\ldots,M_n\, ,
\label{eq:BGT}
\end{multline}
where the scattering phase between $n$- and $m$-strings is
\begin{multline}
  \Theta_{nm}(\lambda) = (1-\delta_{nm})\theta_{|n-m|}(\lambda)+2\theta_{|n-m|+2}(\lambda) \\
+ \ldots + 2\theta_{n+m-2}(\lambda)+\theta_{n+m}(\lambda) \,.
\label{eq:bigthetadef}
\end{multline}

The maximum string quantum number for a given string configuration can now be computed by taking a string center to infinity in the Bethe-Gaudin-Takahashi equations, where the corresponding maximal string quantum number will be given by $I_n^{\max} = I_n^\infty-n$. The number of available states with a string configuration consisting of $M_n$ $n$-strings is then
 \begin{equation}
\prod_n { 2 I_n^{\max}+1 \choose M_n}.
\end{equation}
The number of holes in the sea of $n$-strings is given by $ 2 I_n^{\max}+1 - M_n $.

While for $\Delta=1$ the two-spinon states are formed by all possible sets of real rapidities obeying Bethe equations~\eqref{eq:logbetheeqs}, the four-spinon states contain a two-string, supplemented with real rapidities. In particular, for four-spinon states at $M=N/2-1$, there are $M_1=N/2-3$ one-strings with maximum string quantum number $I^{1,\max}=N/4$, while there is $M_2=1$ two-string with maximum string quantum number $I^{2,\max}=1$, giving rise to 
\begin{equation}
\dim \mathcal{H}^{\Delta=1}_{4\textrm{sp}}=3{N/2+1 \choose N/2-3}
\label{eq:dim4sp}
\end{equation}
possible four-spinon states. 

For a spinon state with a specific set of string quantum numbers $\{ I_j^{(n)} \}$, the rapidities could be obtained by means of an iterative numerical solving procedure of the BGT-equations~\eqref{eq:BGT}. The momentum and energy are then directly determined as a function of the rapidities. Applying the iterative solving procedure to all possible corresponding combinations of string quantum numbers yields the full spectrum of spinon excitations. With these spinon Bethe states, one would now be in place to evaluate the dynamical structure factor in Eq.~\eqref{eq:defDSF1}, by inserting a resolution of the identity and summing over all matrix elements of the ground state with the spinon states,
\begin{equation}
S^{a\bar a}(q,\omega)=2\pi \sum_\alpha |\langle \textrm{GS} | S^-_q |\alpha \rangle |^2 \delta(\omega+\omega_{\textrm{GS}}-\omega_\alpha). \label{eq:defDSF2}
\end{equation}
The matrix elements of spin operators between Bethe states $\langle \{\mu\} | S^a_j | \{\lambda\} \rangle$ are given by normalized determinant expressions\cite{1999_Kitanine_NPB_554,1981_Gaudin_PRD_23,1982_Korepin_CMP_86,2007_Castro_Alvaredo_JPA_40} in terms of the rapidities of the Bethe states. For Bethe states containing string solutions, reduced determinant expressions\cite{2005_Caux_JSTAT_P09003} in terms of the string centers must be employed to overcome divergencies in the determinants caused by the strings.

By integrating Eq.~\eqref{eq:defDSF1} over energy and momentum, sum rules for all matrix elements can be derived. For the transverse dynamical structure factor one obtains for example
\begin{equation}
\int_0^\infty \frac{d \omega}{2\pi} \frac{1}{N} \sum_q S^{-+}(q,\omega)=\frac{1}{2}-\langle S^z \rangle=\frac{M}{N},
\end{equation}
yielding an important quantitative measure of the saturation of the summations over the matrix elements in Eq~\eqref{eq:defDSF2}. In general, two-spinon and four-spinon matrix elements turn out to attribute for the majority of the sum rule saturation for system sizes of the order of a few hundred.

The presence of two-strings in the four-spinon Bethe states raises the question whether string deviations $\delta^{(2)}_j$ from Eq.~\eqref{eq:defstrings} could be neglected. At zero field, a summation over all two-spinon states supplemented with all four-spinon states (neglecting two-string deviations) overshoots the sum rule, implicating that neglecting all deviations leads to erroneous results. String deviations can be handled more carefully by introducing a real parametrization for the two-string rapidities in terms of the two-string center $\lambda^{(2)}_j$ and two-string deviation $\delta^{(2)}_j$,
\begin{equation}
\lambda^{(2),\pm}_j=\lambda^{(2)}_j \pm i \big(\frac{1}{2}+\delta^{(2)}_j \big),
\end{equation}
and solving Bethe equations~\eqref{eq:logbetheeqs} iteratively in this parametrization.\cite{2007_Hagemans_JPA_40} For small rapidity two-string centers, the deviations are exponentially vanishing in system size, while for two-string centers away from zero, the deviations can take on larger values. 

For small string deviations, the determinant expressions become divergent. The reduced form of the determinant~\cite{2005_Caux_JSTAT_P09003} ensures cancellation of these divergencies, which however comes at the price of neglecting the string deviations. In order to handle both large and small two-string deviations, we adopt an algorithm where reduced expressions in terms of the string center are used whenever the string deviation is small, while for large string deviations the rapidities of the deviated string are inserted directly into the original determinant expressions. 

We compute the two-string deviations for all four-spinon states and compute the matrix elements using the aforementioned scheme, keeping track of the deviations in the determinants whenever $|\delta^{(2)}_j | > 10^{-8}$. With this method, the sum rule of all two-spinon and four-spinon contributions is saturated by $99.98\%$ at $N=100$, implying the absence of erroneous contributions which were present when deviations were neglected.

\section{Spinon dynamics in the anisotropic Heisenberg chain}
\label{sec:trackingspinons}
The following section focusses on the real space-time dynamics of spinons, where in particular we elaborate on the construction of the initial state and show results for the time evolution of the magnetization profile. 

The initial state is constructed by acting with a local spin flip on the ground state,
\begin{align}
| \Psi(0) \rangle &=  \sqrt{2} S^+_{j_0} | \{ \lambda_\text{GS}\}\rangle
= \sum_{\{\lambda\}}  C_{\{\lambda\}}  | \{\lambda\}  \rangle \, ,
\label{eq:spinoncn}
\end{align}
where the coefficients $C_{\{\lambda\}}$ are given by the matrix elements of the spin raising operator on site $j_0$,
\begin{equation}
C_{\{\lambda\}}= \sqrt{2} \langle \{\lambda\} |  S^+_{j_0} | \{\lambda_\text{GS}\}\rangle.
\label{eq:clambda}
\end{equation}
The spectral weights $|C_{\{\lambda\}}|^2$ for each spinon state $|\{\lambda\}\rangle$ correspond to the weights as computed or measured from the transverse dynamical structure factor~\eqref{eq:defDSF1} for inelastic neutron scattering experiments. 

The time dependent wave function is computed using unitary time evolution in the basis of Bethe states, 
\begin{equation}
| \Psi (t) \rangle =e^{-i \hat H t} | \Psi(0) \rangle =    \sum_{\{\lambda\}} e^{-i E_{\{\lambda\}} t}  C_{\{\lambda\}}  | \{\lambda\}  \rangle.
\end{equation}
Therefore, the expectation value of the local magnetization at site $j$ is computed by
\begin{equation}
  \langle S^z_j(t) \rangle = \sum_{\{\lambda\},\{\mu\}} e^{-i(E_{\{\lambda\}}-E_{\{\mu\}})t} C_{\{\lambda\}} C_{\{\mu\}}^\ast \langle \{\mu\} | S^z_j | \{\lambda\} \rangle\,.
\label{eq:timeevo}
\end{equation}
The matrix elements between the spinon states occurring in Eqs.~\eqref{eq:clambda} and~\eqref{eq:timeevo} are computed by determinant expressions from algebraic Bethe ansatz results as elaborated in section~\ref{sec:betheansatz}.

A full double Hilbert space summation in equation~(\ref{eq:timeevo}) is not feasible, nonetheless, two-spinon states (Eq.~\eqref{eq:dim2sp}) carry the majority of the spectral weight. In the thermodynamic limit, the two-spinon contributions carry $72.89\%$ of the spectral weight, while at finite system size at $N=100$ considered in the computations of the current work, already $96.85\%$ is ascribed to the two-spinon states. Four-spinon (Eq.~\eqref{eq:dim4sp}) states can be included to achieve a higher overlap of the computed state with the initial state. As described in section~\ref{sec:betheansatz}, the four-spinon states contain deviated two-string solutions, which can be included explicitly by solving the Bethe equations in the parametrization of the deviations and by employing the appropriate reductions of the determinant expressions. As the full spectrum of four-spinon solutions would induce a double summation of $\mathcal{O}(N^8)$ terms, we only select the largest $|C_{\{\lambda\} }|$ of the four-spinon states to make up for $99.00\%$ in total of the sum rule, which for $N=100$ comes down to the inclusion of the $3640$ most important four-spinon states.

\begin{figure}[h!]
\includegraphics[width=\columnwidth]{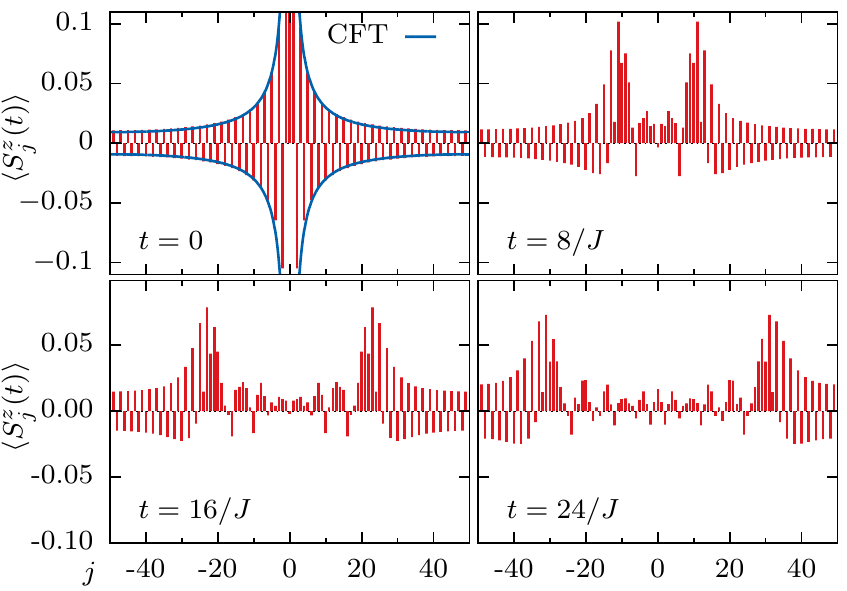}
\caption{Time slices of spinon propagation originating from a local spin flip at $j_0=0$ on the ground state at $N=100$ for $\Delta=1$. The initial state is a projection with total normalization sum rule of $99.00\%$, including all two-spinon states ($96.85\%$) and the most important $3640$ four-spinon states ($2.15\%$). In the top left panel, the CFT prediction from Eq.~\eqref{eq:cft} for the magnetization of the initial state is shown. }
\label{fig:spinonsztimeslicesD1}
\end{figure}

\begin{figure}[h!]
\includegraphics{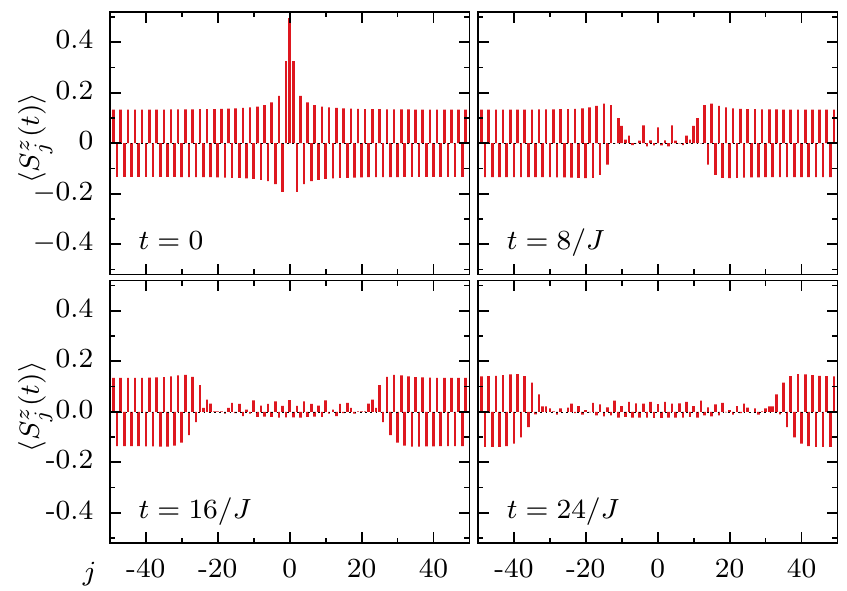}
\caption{Time slices of spinon propagation originating from a spin flip at $j_0=0$ at $N=100$ for $\Delta=1.5$. The initial state is a projection on all two-spinon states with $97.17\%$ saturation of the normalization sum rule.}
\label{fig:spinonsztimeslicesD150}
\end{figure}

\begin{figure}[h!]
\includegraphics{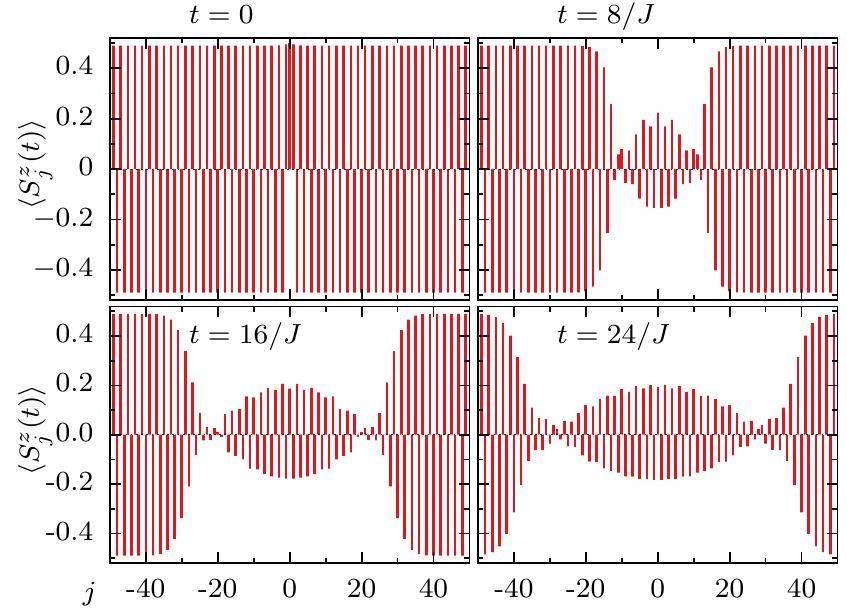}
\caption{Time slices of spinon propagation originating from a spin flip at $j_0=0$ at $N=100$ for $\Delta=10$. The initial state is a projection on all two-spinon states with $99.999\%$ saturation of the normalization sum rule.}
\label{fig:spinonsztimeslicesD10}
\end{figure}

Time slices of the time evolved magnetization profiles evaluated from Eq.~\eqref{eq:timeevo} for propagating spinons are displayed in Figs.~\ref{fig:spinonsztimeslicesD1},~\ref{fig:spinonsztimeslicesD150} and~\ref{fig:spinonsztimeslicesD10} for different values of $\Delta$, including the isotropic case ($\Delta=1$), a case away from the isotropic case and a high value of $\Delta$ towards the Ising limit respectively. In the remainder, we place the up spin at site $j_0=0$.

In the isotropic case $\Delta=1$, the antiferromagnetic ordering in the initial state starts to decrease with increasing distance from the local spin flip. The shape of the magnetization profile of the initial state can be described by the antiferromagnetic longitudinal spin-spin correlations. As the spin at site $j=0$ is always pointing upwards with $S^z_{0}|\Psi(0)\rangle =\frac{1}{2}|\Psi(0)\rangle$, we can identify $\langle S^z_j S^z_0 \rangle = \frac{1}{2}\langle S^z_j \rangle$.  Furthermore, the antiferromagnetic longitudinal spin-spin correlation for the isotropic Heisenberg chain of infinite length can be obtained by the conformal field theory description of the critical low-energy sector,\cite{1986_Affleck_NPB_265,1986_Affleck_PRL_56_1,1987_Affleck_PRB_36,GiamarchiBOOK} and is given to first order as
\begin{equation}
\langle S^z_j S^z_0 \rangle \sim D_1 \frac{(-1)^j}{j}.
\label{eq:cft}
\end{equation}
The prefactor $D_1$ could be fitted from the finite size scaling behavior of the Umklapp matrix elements. \cite{2009_Kitanine_JMP_50,2009_Kitanine_JSTAT_P04003,2011_Shashi_PRB_84,2011_Kitanine_JSTAT_P12010}
Subsequently a conformal transformation to finite size 
\begin{equation}
j \rightarrow \frac{N}{\pi} \sin (j \pi / N)
\label{eq:conftransfinitesize}
\end{equation}
 is applied. In Fig.~\ref{fig:spinonsztimeslicesD1}, the conformal field theory prediction of the initial magnetization profile of the spinon state from Eq.~\eqref{eq:cft} is plotted on top of the algebraic Bethe ansatz finite size computation of the magnetization, showing agreement between the results.

At $\Delta \rightarrow \infty$ the ground state $| \{\lambda\}_\text{GS}\rangle$ becomes equal to the Ne\'el state,
\begin{equation}
| \textrm{Ne\'el} \rangle = \frac{1}{\sqrt{2}} \left( | \uparrow \downarrow \cdot\cdot\cdot \rangle + | \downarrow \uparrow  \cdot\cdot\cdot \rangle \right)\, ,
\label{eq:neelstate}
\end{equation}
so the spin raising operator creates exactly three adjacent spins aligned upwards in the antiferromagnetic ordering. By acting with the spin raising operator, only one of the two antiferromagnetic ordered states in Eq.~\eqref{eq:neelstate} is selected. The state with high $\Delta$ shown in Fig.~\ref{fig:spinonsztimeslicesD10} has nearly perfect N\'eel order of the spins at $\langle S^z_{j}\rangle=\pm \frac{1}{2}$.  Furthermore, at $\Delta \rightarrow \infty$, the initial state is fully overlapped with two-spinon states only, making the computation of observables exact in this limit.

From Fig.~\ref{fig:spinonsztimeslicesD10} a few observations could be made. The spinon is moving like a domain wall, as all spins flip to their opposite signs after the passing of the spinons. However, this domain wall is not strictly localized around one lattice site, but rather has a finite extent.  After the spinon passed by, the magnetization does not restore to exactly the opposite of the initial value as one would expect from the propagation of a pure, strictly localized, single domain wall between the two Ne\'el ordered domains, but remains around $0.2$. This behavior remains even for much higher values of $\Delta$.  Furthermore, the expectation value $\langle S^z_0 \rangle$ of the spin at $j=0$ immediately starts to decrease rather than staying pointing upwards, suggesting that the two domain walls created around $j=0$ start to propagate in both directions, rather than moving only outwards with respect to $j=0$ (which should have let the spin at site $j=0$ pointing upwards).

\begin{figure}[h!]
\includegraphics{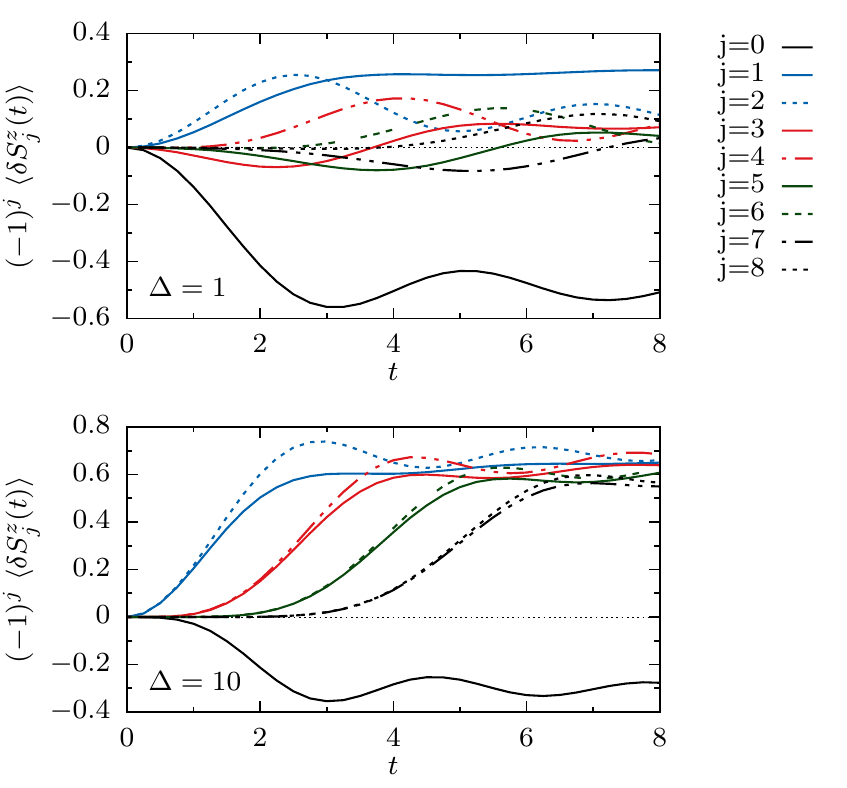}
\caption{Spinon propagation shown in plots of the magnetic fluctuations $\langle \delta S^z_j(t)\rangle$ at single sites as a function of time for $\Delta=1$ (top) and $\Delta=10$ (bottom).}
\label{fig:spinonsinglesite}
\end{figure}

Further behavior of the spin flip on the antiferromagnetic ground state could be studied from the fluctuations of the spin expectation values from the initial state, which would also be useful for the study of correlations in the next section,
 \begin{equation}
\delta S^z_j(t) = S^z_j(t)-S^z_j(0)\, .
\label{eq:deffluc}
\end{equation}

Single site plots for the fluctuations from the initial state for the magnetization $\langle \delta S^z_j(t) \rangle$ are shown in Fig.~\ref{fig:spinonsinglesite}. For the $\Delta=10$ case, it is noted that two adjacent spins anti-aligned originally, start collapsing almost simultaneously as the spinon passes by.

\section{Spinon dynamics in the Babujan-Takhtajan spin-1 chain}
\label{sec:BT}
The availability of determinant expressions for matrix elements of local spin operators for integrable higher spin chains\cite{2007_Castro_Alvaredo_JPA_40} allows for an extension of the time-evolution computation after a local spin flip towards the Babujan-Takhtajan spin-1 chain, of which the Hamiltonian reads\cite{1981_Kulish_LMP_5,1982_Takhtajan_PLA_87,1982_Babujan_PLA_90,1983_Babujan_NPB_215}
\begin{equation}
H=\frac{J}{4} \sum_j \left[  \hat  S_{j}\hat S_{j+1} - (\hat S_{j} \hat S_{j+1})^{2} \right].
\end{equation}
The dynamical stucture factor of this integrable spin-$1$ chain has been computed as well, based on summations of matrix elements of the two-spinon and four-spinon excitations.\cite{2014_Vlijm_JSTAT_P05009}

The corresponding Bethe equations of this integrable higher spin chain are similar to the spin-$\frac{1}{2}$ isotropic Heisenberg model, with only a modification in the first term with respect to Eq.~\eqref{eq:logbetheeqs},
\begin{equation}
\theta_2(\lambda_j) - \frac{1}{N} \sum_{k \neq j}^M \theta_2(\lambda_j-\lambda_k) = \frac{2 \pi}{N} J_j,
\label{eq:logbetheeqss1}
\end{equation}
where the functions $\theta_n(\lambda)$ are given by Eq.~\eqref{eq:thetadef1}. 

The ground state for $J>0$ favors antiferromagnetic ordering with $M=N$ down spins (with zero total magnetization), for which the corresponding set of rapidities consists of $M_2=N/2$ two-strings. The existence of a macroscopic number of bound states already in the ground state (as opposed to spin-1/2 models), makes the analysis of two-string deviations even more important. To leading order in system size, the two-string deviations for the ground state are of the form\cite{1990_deVega_JPA_23,1990_Kluemper_JPA_23} $\delta_j^{(2)} \approx \ln(2) \cosh(\pi \lambda)/(2\pi N)$, implying that all two-string deviations are only algebraically small in system size.

The most important excitations after a spin flip can be considered by breaking up one of the two-strings from the ground state, and placing one rapidity back on the real axis. The number of available Bethe states of this type of excitations with string configuration $M_1=1$, $M_2=N/2-1$ is determined from the limiting quantum numbers and turns out to be the same as Eq.~\eqref{eq:dim2sp} for the two-spinon excitations of the isotropic spin-$\frac{1}{2}$ model. For the spin-1 model, this type of excitations could be pictured as two hole-like modes in the sea of two-strings, being similar to two hole-like modes in the sea of one-strings resembling two-spinon states in the spin-$\frac{1}{2}$ chain. Four-spinons in the spin-$1$ chain could be considered by adding a three-string, but are left out of consideration in this analysis.

\begin{figure}[h!]
\includegraphics{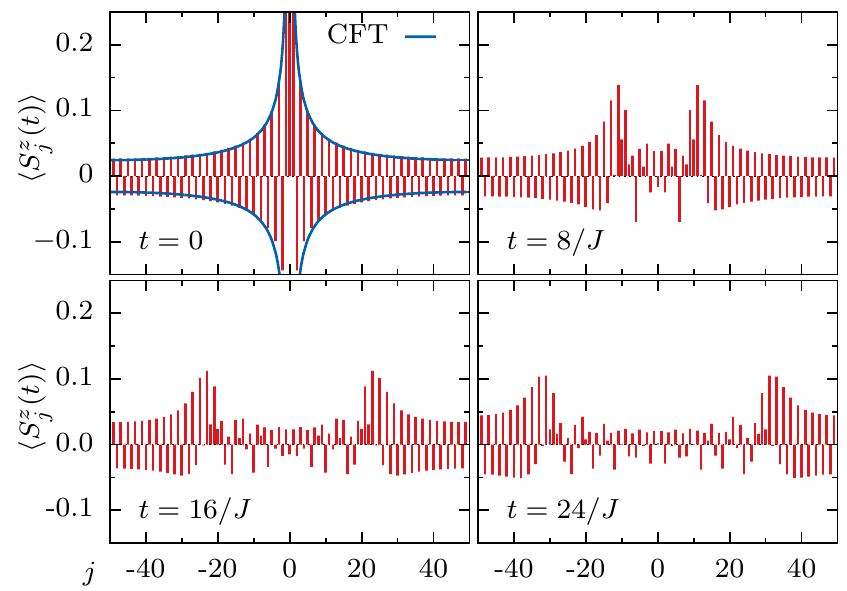}
\caption{Time slices of two-spinon propagation originating from a spin flip at $j_0=0$ at $N=100$ for the isotropic spin-1 Babujan-Takhatajan model. The initial state is a projection on all two-spinon states (consisting of hole-like modes in a sea of two-strings) with $92.33\%$ saturation of the normalization sum rule.}
\label{fig:spinonsztimeslicesS1}
\end{figure}

Fig.~\ref{fig:spinonsztimeslicesS1} shows the time evolution of a local spin flip on the ground state of the Babujan-Takhtajan spin-1 chain, projected on all two-spinon states in the $M=N-1$ sector, for a chain consisting of $N=100$ sites. The sum rule saturation of all two-spinon states is $92.33\%$ at this system size.  All string deviations have been included in the computation of the relevant matrix elements. 

The initial shape of the local magnetization can again be extracted from conformal field theory. The critical low-energy sector is described by the SU($2$) level-$2$ Wess-Zumino-Novikov-Witten model\cite{1988_Alcaraz_JPA_21,1990_Tsvelik_PRB_42_16,2013_Michaud_PRB_87,1984_Knizhnik_NPB_247}, which yields for the asymptotic behavior of the longitudinal antiferromagnetic spin-spin correlations of an infinite chain
\begin{equation}
\langle S^z_j S^z_0 \rangle \sim D_1 \frac{(-1)^j}{j^\frac{3}{4}}.
\label{eq:cfts1}
\end{equation}
With the appropriate transformation to finite size (see Eq.~\eqref{eq:conftransfinitesize}), Eq.~\eqref{eq:cfts1} is plotted on top of the computed data in Fig.~\ref{fig:spinonsztimeslicesS1}, showing agreement for the shape of the initial magnetization profile for both approaches. The slight discrepancy is attributed to the incomplete saturation of the normalization sum rule.

\section{Spin-spin correlations}
\label{sec:correlations}
Besides the magnetization expectation value, the method of summing over matrix elements of local spin operators between Bethe states at finite size allows for the computation of two-point correlations by inserting a resolution of the identity and summing over all relevant matrix elements. In this way, one can for example display the buildup of correlations between two separated spins after a spin flip is applied in the middle of the two spins. Moreover, the nearest-neighbor spin-spin correlation could be studied as well while the spinons pass by at a specific site. Both quantities acquire an error in the results due to the usage of an extra (truncated) summation over states, which could be quantified by keeping track of sum rules and by comparing the results for low system sizes to exact numerical diagonalization. We restrict to the isotropic Heisenberg model ($\Delta=1$) in the analysis of spin-spin correlations in this section.

The equal time spin-spin correlation between two spins at sites $\pm j$ yields, see also Fig.~\ref{fig:spinsforcorrelation} for illustration,
\begin{equation}
\langle \Psi(0) |  \delta S^z_j(t) \delta S^z_{-j}(t) | \Psi(0) \rangle,
\label{eq:2pjminj}
\end{equation}
while the spin-flip creating the spinons is located in the middle of the two spins at $j=0$ in the initial state $| \Psi(0) \rangle$. The consideration of the fluctuations of $S^z_j$ with respect to the initial state (see Eq.~\eqref{eq:deffluc}) accounts for the initial values of the spin-spin correlation and assures that Eq.~\eqref{eq:2pjminj} starts from zero at zero time. 

\begin{figure}[h!]
\begin{align*}
S_0^+ \qquad \qquad \;\;\;\;\;\, \\
\cdot\cdot\cdot \downarrow \; \uparrow \; \downarrow \; \uparrow \; \downarrow \; \uparrow \; \uparrow \; \uparrow \; \downarrow \; \uparrow \; \downarrow \; \uparrow \;\downarrow \cdot\cdot\cdot \\
\delta S_{-j}^{z}(t) \qquad \qquad \qquad  \delta S_j^z(t)
\end{align*}
\caption{Illustration of the location of the spins at sites $\pm j$ between which the correlation is computed in Eq.~\eqref{eq:2pjminj}, relative to the location of the spin flip creating the spinon excitations.}
\label{fig:spinsforcorrelation}
\end{figure}

\begin{figure}[h!]
\includegraphics{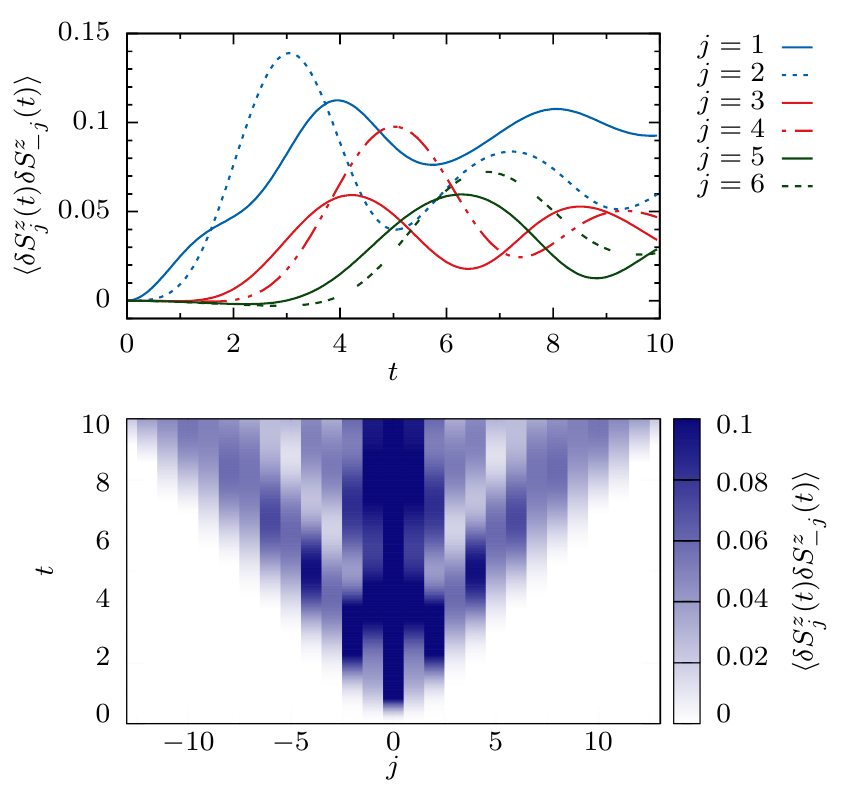}
\caption{The spin-spin correlation between two sites at $\pm j$, with the up spin is in the middle at $j=0$, computed for $\Delta=1$ and $N=32$. The projection of the initial state on two-spinon states saturates $98.82\%$ of the normalization sum rule. Top: single site plots as function of time. Bottom: heatmap showing the light-cone effect in the build up of the correlations at the spinon propagation velocity.}
\label{fig:corr1}
\end{figure}

Fig.~\ref{fig:corr1} shows the time evolved results of the computation of Eq.~\eqref{eq:2pjminj}. As the spinon passes by at site $j$, it becomes correlated to the spin at site $-j$ opposite with respect to the spin flip. The propagation of the build up of correlations takes place at the spinon propagation velocity $v=J \pi /2$, giving a clear image of a light-cone effect, with zero correlation outside the light cone. Applying a local spin flip on the ground state therefore satisfies the Lieb-Robinson bound\cite{1972_Lieb_CMP_28}, stating that no correlation can propagate faster than the (spinon) velocity in the system.

\begin{figure}[h!]
\includegraphics{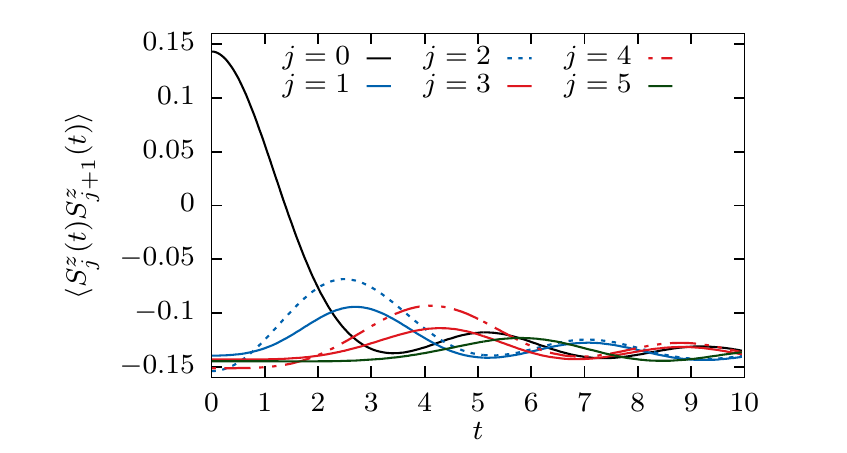}
\caption{Longitudinal nearest-neighbor correlation at $\Delta=1$ and $N=32$. The projection on two-spinon states saturates $98.82\%$ of the normalization sum rule. As the spinon passes by, the antiferromagnetic correlation becomes weaker and then revives again, showing oscillations after the passage of the spinon.  }
\label{fig:szsznn}
\end{figure}

The antiferromagnetic nearest-neighbor correlations are computed from
\begin{equation}
\langle \Psi(0) |  S^z_j(t) S^z_{j+1}(t) | \Psi(0) \rangle,
\label{eq:2pnn}
\end{equation}
while the results of this computation (again by inserting a resolution of the identity and summing over matrix elements) are plotted in Fig.~\ref{fig:szsznn}. As the spinon passes by, the antiferromagnetic correlation collapses and then revives again, showing oscillations after the passage of the spinon. This observation is in particular interesting as the expectation value of the local magnetization collapses at $\Delta=1$ (see Fig.~\ref{fig:spinonsztimeslicesD1}), while the antiferromagnetic correlation is restored after the passage of the spinon.

\begin{figure}[h!]
\includegraphics{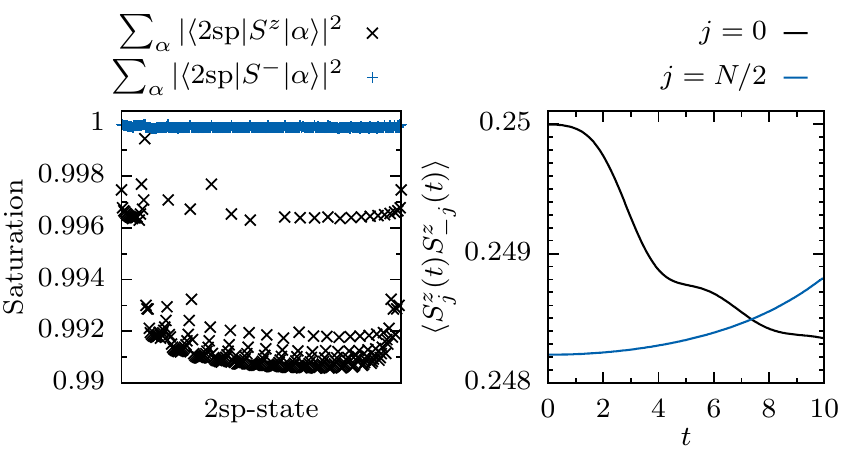}
\caption{Errors of the method induced by the truncation of the sum over intermediate states at $\Delta=1$. Left: Sum rule saturation for every two-spinon state, by summing over the intermediate states with zero and one infinite rapidity respectively. Right: Expectation value of longitudinal spin-spin correlation on the same site, which should be $1/4$.}
\label{fig:correrror}
\end{figure}

The computational method adopted in this section for the spin-spin correlations requires an extra resolution of the identity $1=\sum_{\{\alpha\}} | \{ \alpha \} \rangle \langle \{ \alpha \} |$ between the two spin operators in the two-point functions of Eqs.~\eqref{eq:2pjminj} and~\eqref{eq:2pnn}. For the matrix elements of $S_z$ for the isotropic case $\Delta=1$, the intermediate states $|\{\alpha\}\rangle$ can contain either zero or one infinite rapidity. These matrix elements of states containing an infinite rapidity can be expressed as matrix elements of states without infinite rapidities, \cite{1981_Mueller_PRB_24}
\begin{equation}
\langle \{ \mu \}_M | S^z | \{ \lambda, \infty \}_M \rangle = -\frac{1}{2} \langle \{ \mu \}_M | S^- | \{ \lambda \}_{M-1} \rangle.
\end{equation}
Again, it is computationally unfeasible to include all states in the summation over intermediate states. We therefore restrict to the highest-weight two-spinon and four-spinon states at $M=N/2-1$ as constructed in Sec II, and the lower weight states containing $M_1=N/2-2$ real rapidities and one infinite rapidity, and the lower weight states with $M_1=N/2-4$, $M_2=1$ and one infinite rapidity. For the latter, two-string deviations are properly taken into account. The errors induced by the truncation of the summation to these types of intermediate states are quantified in Fig.~\ref{fig:correrror}. The saturation of the extra intermediate summations on all two-spinon and four-spinon states exceeds at least $99\%$, while for known exact values for the spin-spin correlation at $j=0$, the induced error stays within $1\%$ of the exact result.

\begin{figure}[h!]
\includegraphics{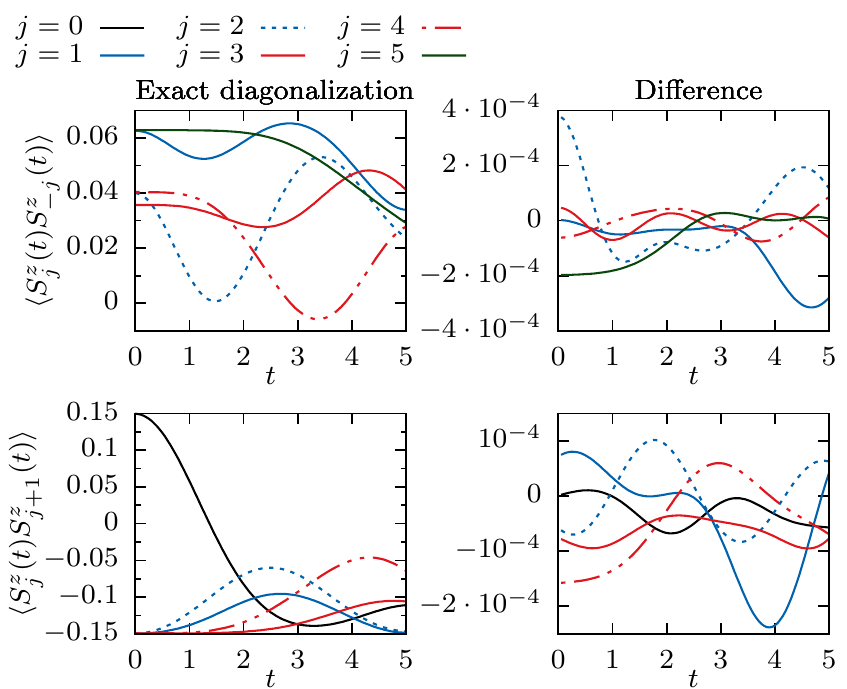}
\caption{Left panels: Exact diagonalization results for $N=12$ for the spin-spin correlations from Eq.~\eqref{eq:2pjminj} (top) and the nearest neighbor correlations from Eq.~\eqref{eq:2pnn} (bottom). Right panels: difference between the results from exact diagonalization and matrix element summations from algebraic Bethe ansatz.}
\label{fig:correrrorexactdiag}
\end{figure}

An additional reliability check of the computations of spin-spin correlations based on matrix elements of single spin operators (including all two-spinon and four-spinon states) can be performed by comparison with exact diagonalization at low system size ($N=12$). From the right panels of Fig~\ref{fig:correrrorexactdiag}, it can be seen that the differences with the results from exact diagonalization are small.  

\section{Conclusions}\label{sec:conclusions}
We demonstrated the real space-time dynamics of spinons in the (an)isotropic Heisenberg chain and the Babujan-Takhtajan spin-1 chain, initiated from a local spin flip on the antiferromagnetic ground state, reflecting precisely the state for which the spectral weights of the spinon states are measured or computed in the dynamical structure factor for inelastic neutron scattering purposes. 

The propagation of the domain walls in the antiferromagnetic ordering of the spins is visualized for different values of anisotropy by computing the local expectation value of the magnetization as a function of time. This method is based on the algebraic Bethe ansatz for the Heisenberg model at finite system size. By an iterative procedure, it is possible to solve the spinon Bethe states explicitly. Subsequently, matrix element expressions in determinant form for local spin operators between the Bethe states allow to construct the initial state with a local spin flip and to compute the time evolution of its local magnetization and spin-spin correlations.

The magnetization profiles for the propagating spinons show almost simultaneous spin flip processes for pairs of neighboring anti-aligned spins. After the passing of a spinon, the antiferromagnetic order is destroyed for $\Delta=1$, but the antiferromagnetic correlation between nearest neighbors survives, showing oscillatory behavior. At high $\Delta$ towards the Ising limit, no revival of the magnetization to its initial value is visible, while the domain wall is not sharp either. These observations can be explained from the domain walls created adjacent to the initial spin flip, which can propagate in both directions simultaneously after the spin flip. Moreover, correlations between two separated spins were considered, showing the buildup of correlations only after the passing of the spinon, displaying a light-cone effect. The evolution of the correlation therefore satisfies the Lieb-Robinson bound, as the correlations spread at the spinon propagation velocity. 

The algebraic Bethe ansatz methods at finite size employed here could be implemented towards more out-of-equilibrium situations in spin chains, for which the dynamics could be computed from the determinant expressions for matrix elements of spin operators between Bethe states. Candidates for future applications of this method are the addition of non-integrable spin-spin interactions or driven magnetic fields to the Heisenberg Hamiltonian. 

\section*{Acknowledgements} 
The authors acknowledge support from the Foundation for Fundamental Research on Matter (FOM) and from the Netherlands Organization for Scientific Research (NWO). This work forms
part of the activities of the Delta-Institute for Theoretical Physics (D-ITP).


\bibliographystyle{unsrt}

\begin{thebibliography}{10}

\bibitem{1981_Faddeev_PLA_85}
L.~D. Faddeev and L.~A. Takhtajan.
\newblock What is the spin of a spin wave?
\newblock {\em Phys. Lett. A}, 85:375, 1981.

\bibitem{1928_Heisenberg_ZP_49}
W.~Heisenberg.
\newblock {Zur Theorie des Ferromagnetismus}.
\newblock {\em Z. Phys.}, 49:619, 1928.

\bibitem{1931_Bethe_ZP_71}
H.~Bethe.
\newblock Zur {T}heorie der {M}etalle. i. {E}igenwerte und {E}igenfunktionen
  der linearen {A}tomkette.
\newblock {\em Zeit. f\"ur Physik}, 71:205, 1931.

\bibitem{1958_Orbach_PR_112}
R.~Orbach.
\newblock Linear antiferromagnetic chain with anisotropic coupling.
\newblock {\em Phys. Rev.}, 112(2):309--316, 1958.

\bibitem{KorepinBOOK}
V.~E. Korepin, N.~M. Bogoliubov, and A.~G. Izergin.
\newblock {\em Quantum Inverse Scattering Method and Correlation Functions}.
\newblock Cambridge Univ. Press, 1993.

\bibitem{1991_Nagler_PRB_44}
S.~E. Nagler, D.~A. Tennant, R.~A. Cowley, T.~G. Perring, and S.~K. Satija.
\newblock {Spin dynamics in the quantum antiferromagnetic chain compound
  ${\mathrm{KCuF}}_{3}$}.
\newblock {\em Phys. Rev. B}, 44:12361--12368, 1991.

\bibitem{1993_Tennant_PRL_70}
D.~A. Tennant, T.~G. Perring, R.~A. Cowley, and S.~E. Nagler.
\newblock {Unbound spinons in the S=1/2 antiferromagnetic chain $KCuF3$}.
\newblock {\em Phys. Rev. Lett.}, 70(25):4003--4006, 1993.

\bibitem{1995_Tennant_PRB_52_1}
D.~A. Tennant, S.~E. Nagler, D.~Welz, G.~Shirane, and K.~Yamada.
\newblock {Effects of coupling between chains on the magnetic excitation
  spectrum of $KCuF3$}.
\newblock {\em Phys. Rev. B}, 52(18):13381--13389, 1995.

\bibitem{1995_Tennant_PRB_52_2}
D.~A. Tennant, R.~A. Cowley, S.~E. Nagler, and A.~M. Tsvelik.
\newblock {Measurement of the spin-excitation continuum in one-dimensional
  $KCuF_{3}$ using neutron scattering}.
\newblock {\em Phys. Rev. B}, 52(18):13368--13380, 1995.

\bibitem{2004_Zaliznyak_PRL_93}
I.~A. Zaliznyak, H.~Woo, T.~G. Perring, C.~L. Broholm, C.~D. Frost, and
  H.~Takagi.
\newblock {Spinons in the Strongly Correlated Copper Oxide Chains in $SrCuO2$}.
\newblock {\em Phys. Rev. Lett.}, 93(8):087202, 2004.

\bibitem{2013_Mourigal_NATPHYS_9}
M.~Mourigal, M.~Enderle, A.~Kl{\"o}pperpieper, J.-S. Caux, A.~Stunault, and
  H.~M. R{\o}nnow.
\newblock Fractional spinon excitations in the quantum heisenberg
  antiferromagnetic chain.
\newblock {\em Nat. Phys.}, 9(7):435 -- 441, 2013.

\bibitem{2013_Lake_PRL_111}
B.~Lake, D.~A. Tennant, J.-S. Caux, T.~Barthel, U.~Schollw\"ock, S.~E. Nagler,
  and C.~D. Frost.
\newblock {Multispinon Continua at Zero and Finite Temperature in a Near-Ideal
  Heisenberg Chain}.
\newblock {\em Phys. Rev. Lett.}, 111:137205, 2013.

\bibitem{1954_vanHove_PR_95_1}
L.~Van~Hove.
\newblock Correlations in space and time and born approximation scattering in
  systems of interacting particles.
\newblock {\em Phys. Rev.}, 95(1):249--262, 1954.

\bibitem{1954_vanHove_PR_95_2}
L.~Van~Hove.
\newblock Time-dependent correlations between spins and neutron scattering in
  ferromagnetic crystals.
\newblock {\em Phys. Rev.}, 95(6):1374--1384, 1954.

\bibitem{1989_Slavnov_TMP_79}
N.~A. Slavnov.
\newblock Calculation of scalar products of wave functions and form factors in
  the framework of the algebraic {B}ethe {A}nsatz.
\newblock {\em Theor. Math. Phys.}, 79:502, 1989.

\bibitem{1990_Slavnov_TMP_82}
N.~A. Slavnov.
\newblock Nonequal-time current correlation function in a one-dimensional bose
  gas.
\newblock {\em Theor. Math. Phys.}, 82:273, 1990.

\bibitem{1999_Kitanine_NPB_554}
N.~Kitanine, J.~M. Maillet, and V.~Terras.
\newblock Form factors of the {XXZ} {H}eisenberg finite chain.
\newblock {\em Nucl. Phys. B}, 554(3):647 -- 678, 1999.

\bibitem{2007_Castro_Alvaredo_JPA_40}
O.~A. Castro-Alvaredo and J.~M. Maillet.
\newblock {Form factors of integrable Heisenberg (higher) spin chains}.
\newblock {\em J. Phys. A: Math. Theor.}, 40(27):7451, 2007.

\bibitem{2002_Biegel_EPL_59}
D.~Biegel, M.~Karbach, and G.~M{\"u}ller.
\newblock {Transition rates via Bethe ansatz for the spin-1/2 Heisenberg
  chain}.
\newblock {\em EPL (Europhysics Letters)}, 59(6):882, 2002.

\bibitem{2003_Biegel_JPA_36}
D~Biegel, M~Karbach, and G~M{\"u}ller.
\newblock Transition rates via bethe ansatz for the spin-1/2 planar xxz
  antiferromagnet.
\newblock {\em Journal of Physics A: Mathematical and General},
  36(20):5361--5368, 2003.

\bibitem{2004_Sato_JPSJ_73}
J.~Sato, M.~Shiroishi, and M.~Takahashi.
\newblock {Evaluation of Dynamic Spin Structure Factor for the Spin-$1/2$ XXZ
  Chain in a Magnetic Field}.
\newblock {\em Jour. Phys. Soc. Jpn}, 73(11):3008--3014, 2004.

\bibitem{2005_Caux_JSTAT_P09003}
J.-S. Caux, R.~Hagemans, and J.~M. Maillet.
\newblock {Computation of dynamical correlation functions of Heisenberg chains:
  the gapless anisotropic regime}.
\newblock {\em J. Stat. Mech.: Th. Exp.}, 2005(09):P09003, 2005.

\bibitem{2005_Caux_PRL_95}
J.-S. Caux and J.~M. Maillet.
\newblock {Computation of Dynamical Correlation Functions of Heisenberg Chains
  in a Magnetic Field}.
\newblock {\em Phys. Rev. Lett.}, 95(7):077201, 2005.

\bibitem{2009_Kohno_PRL_102}
M.~Kohno.
\newblock {Dynamically Dominant Excitations of String Solutions in the
  Spin-$1/2$ Antiferromagnetic Heisenberg Chain in a Magnetic Field}.
\newblock {\em Phys. Rev. Lett.}, 102(3):037203, 2009.

\bibitem{JimboBOOK}
M.~Jimbo and T.~Miwa.
\newblock {\em {Algebraic Analysis of Solvable Lattice Models}}.
\newblock American Mathematical Society, Providence, RI, 1995.

\bibitem{1996_Bougourzi_PRB_54}
A.~H. Bougourzi, M.~Couture, and M.~Kacir.
\newblock {Exact two-spinon dynamical correlation function of the
  one-dimensional Heisenberg model}.
\newblock {\em Phys. Rev. B}, 54(18):R12669--R12672, 1996.

\bibitem{1997_Karbach_PRB_55}
M.~Karbach, G.~M\"uller, A.~H. Bougourzi, A.~Fledderjohann, and K.-H. M\"utter.
\newblock {Two-spinon dynamic structure factor of the one-dimensional $s=1/2$
  Heisenberg antiferromagnet}.
\newblock {\em Phys. Rev. B}, 55(18):12510--12517, 1997.

\bibitem{1998_Bougourzi_PRB_57}
A.~Hamid Bougourzi, Michael Karbach, and Gerhard M\"uller.
\newblock {Exact two-spinon dynamic structure factor of the one-dimensional
  $s=12$ Heisenberg-Ising antiferromagnet}.
\newblock {\em Phys. Rev. B}, 57(18):11429--11438, 1998.

\bibitem{2008_Caux_JSTAT_P08006}
J.-S. Caux, J.~Mossel, and I.~P{\'e}rez Castillo.
\newblock {The two-spinon transverse structure factor of the gapped Heisenberg
  antiferromagnetic chain}.
\newblock {\em J. Stat. Mech.: Th. Exp.}, 2008(08):P08006, 2008.

\bibitem{1997_Abada_NPB_497}
A.~Abada, A.~H. Bougourzi, and B.~Si-Lakhal.
\newblock {Exact four-spinon dynamical correlation function of the Heisenberg
  model}.
\newblock {\em Nucl. Phys. B}, 497(3):733 -- 753, 1997.

\bibitem{2006_Caux_JSTAT_P12013}
J.-S. Caux and R.~Hagemans.
\newblock {The four-spinon dynamical structure factor of the Heisenberg chain}.
\newblock {\em J. Stat. Mech: Th. Exp.}, 2006(12):P12013, 2006.

\bibitem{2015_Deguchi_ArXiv}
T.~Deguchi, P.~R. Giri, and R.~Hatakeyama.
\newblock { Power-law relaxation behavior of an initially localized state in
  the spin-1/2 Heisenberg chain }.
\newblock {\em ArXiv:1507.07470}, 2015.

\bibitem{1972_Lieb_CMP_28}
E.~H. Lieb and D.~W. Robinson.
\newblock The finite group velocity of quantum spin systems.
\newblock {\em Commun. Math. Phys.}, 28(3):251--257, 1972.

\bibitem{GaudinBOOK}
M.~Gaudin.
\newblock {\em La fonction d'onde de {B}ethe}.
\newblock Masson, Paris, 1983.

\bibitem{GaudinTRANSLATION}
M.~Gaudin.
\newblock {\em The Bethe Wavefunction (Translated by J.-S. Caux)}.
\newblock Cambridge University Press, 2014.

\bibitem{1972_Takahashi_PTP_48}
M.~Takahashi and M.~Suzuki.
\newblock {One-Dimensional Anisotropic Heisenberg Model at Finite
  Temperatures}.
\newblock {\em Prog. Theor. Phys.}, 48(6):2187--2209, 1972.

\bibitem{1981_Gaudin_PRD_23}
M.~Gaudin, B.~M. McCoy, and T.~T. Wu.
\newblock {Normalization sum for the Bethe's hypothesis wave functions of the
  Heisenberg-Ising chain}.
\newblock {\em Phys. Rev. D}, 23(2):417--419, 1981.

\bibitem{1982_Korepin_CMP_86}
V.~E. Korepin.
\newblock {Calculation of norms of Bethe wave functions}.
\newblock {\em Commun. Math. Phys.}, 86:391--418, 1982.

\bibitem{2007_Hagemans_JPA_40}
R.~Hagemans and J.-S. Caux.
\newblock Deformed strings in the {H}eisenberg model.
\newblock {\em J. Phys. A: Math. Theor.}, 40(49):14605, 2007.

\bibitem{1986_Affleck_NPB_265}
I.~Affleck.
\newblock {Exact critical exponents for quantum spin chains, non-linear
  $\sigma$-models at $\theta = \pi$ and the quantum Hall effect}.
\newblock {\em Nucl. Phys. B}, 265(3):409 -- 447, 1986.

\bibitem{1986_Affleck_PRL_56_1}
I.~Affleck.
\newblock Universal term in the free energy at a critical point and the
  conformal anomaly.
\newblock {\em Phys. Rev. Lett.}, 56:746--748, 1986.

\bibitem{1987_Affleck_PRB_36}
I.~Affleck and F.~D.~M. Haldane.
\newblock Critical theory of quantum spin chains.
\newblock {\em Phys. Rev. B}, 36:5291--5300, 1987.

\bibitem{GiamarchiBOOK}
T.~Giamarchi.
\newblock {\em Quantum Physics in One Dimension}.
\newblock Oxford University Press, 2004.

\bibitem{2009_Kitanine_JMP_50}
N.~Kitanine, K.~K. Kozlowski, J.~M. Maillet, N.~A. Slavnov, and V.~Terras.
\newblock {On the thermodynamic limit of form factors in the massless XXZ
  Heisenberg chain}.
\newblock {\em J. Math. Phys.}, 50(9):095209, 2009.

\bibitem{2009_Kitanine_JSTAT_P04003}
N.~Kitanine, K.~K. Kozlowski, J.~M. Maillet, N.~A. Slavnov, and V.~Terras.
\newblock {Algebraic Bethe ansatz approach to the asymptotic behavior of
  correlation functions}.
\newblock {\em J. Stat. Mech.: Th. Exp.}, 2009(04):P04003, 2009.

\bibitem{2011_Shashi_PRB_84}
A.~Shashi, L.~I. Glazman, J.-S. Caux, and A.~Imambekov.
\newblock Nonuniversal prefactors in the correlation functions of
  one-dimensional quantum liquids.
\newblock {\em Phys. Rev. B}, 84(4):045408, 2011.

\bibitem{2011_Kitanine_JSTAT_P12010}
N.~Kitanine, K.~K. Kozlowski, J.~M. Maillet, N.~A. Slavnov, and V.~Terras.
\newblock A form factor approach to the asymptotic behavior of correlation
  functions in critical models.
\newblock {\em Journal of Statistical Mechanics: Theory and Experiment},
  2011(12):P12010, 2011.

\bibitem{1981_Kulish_LMP_5}
P.~Kulish, N.~Reshetikhin, and E.~Sklyanin.
\newblock {Yang-Baxter equation and representation theory: I}.
\newblock {\em Lett. Math. Phys.}, 5:393--403, 1981.

\bibitem{1982_Takhtajan_PLA_87}
L.A. Takhtajan.
\newblock {The picture of low-lying excitations in the isotropic Heisenberg
  chain of arbitrary spins}.
\newblock {\em Phys. Lett. A}, 87(9):479 -- 482, 1982.

\bibitem{1982_Babujan_PLA_90}
H.M. Babujan.
\newblock {Exact solution of the one-dimensional isotropic Heisenberg chain
  with arbitrary spins S}.
\newblock {\em Phys. Lett. A}, 90(9):479 -- 482, 1982.

\bibitem{1983_Babujan_NPB_215}
H.M. Babujan.
\newblock {Exact solution of the isotropic Heisenberg chain with arbitrary
  spins: Thermodynamics of the model}.
\newblock {\em Nucl. Phys. B}, 215(3):317 -- 336, 1983.

\bibitem{2014_Vlijm_JSTAT_P05009}
R.~Vlijm and J.-S. Caux.
\newblock { Computation of dynamical correlation functions of the spin-1
  Babujan–Takhtajan chain}.
\newblock {\em JSTAT}, page P05009, 2014.

\bibitem{1990_deVega_JPA_23}
H.~J. de~Vega and F.~Woynarovich.
\newblock {Solution of the Bethe Ansatz equations with complex roots for finite
  size: the spin $S=1$ isotropic and anisotropic chains}.
\newblock {\em J. Phys. A: Math. Gen.}, 23(9):1613, 1990.

\bibitem{1990_Kluemper_JPA_23}
A.~Kl{\"u}mper and M.~T. Batchelor.
\newblock {An analytic treatment of finite-size corrections in the spin-1
  antiferromagnetic XXZ chain}.
\newblock {\em J. Phys. A: Math. Gen.}, 23(5):L189, 1990.

\bibitem{1988_Alcaraz_JPA_21}
F.~C. Alcaraz and M.~J. Martins.
\newblock {Conformal anomaly and critical exponents of the spin-1
  Takhtajan-Babujan model}.
\newblock {\em J. Phys. A: Math. Gen.}, 21(7):L381, 1988.

\bibitem{1990_Tsvelik_PRB_42_16}
A.~M. Tsvelik.
\newblock {Field-theory treatment of the Heisenberg spin-1 chain}.
\newblock {\em Phys. Rev. B}, 42:10499--10504, 1990.

\bibitem{2013_Michaud_PRB_87}
F.~Michaud, S.~R. Manmana, and F.~Mila.
\newblock {Realization of higher Wess-Zumino-Witten models in spin chains}.
\newblock {\em Phys. Rev. B}, 87:140404, 2013.

\bibitem{1984_Knizhnik_NPB_247}
V.G. Knizhnik and A.B. Zamolodchikov.
\newblock {Current algebra and Wess-Zumino model in two dimensions}.
\newblock {\em Nucl. Phys. B}, 247(1):83 -- 103, 1984.

\bibitem{1981_Mueller_PRB_24}
G.~M\"uller, H.~Thomas, H.~Beck, and J.~C. Bonner.
\newblock Quantum spin dynamics of the antiferromagnetic linear chain in zero
  and nonzero magnetic field.
\newblock {\em Phys. Rev. B}, 24(3):1429--1467, 1981.

\end{thebibliography}
\addcontentsline{toc}{chapter}{Bibliography}

\end{document}